\documentstyle[12pt]{article}

\begin{document}
\input{psfig.sty}
\begin{flushright}
\baselineskip=12pt
UPR-954-T \\
\end{flushright}

\begin{center}
\vglue 1.5cm
{\Large\bf Local Discrete Symmetry in the Brane Neighborhood}
\vglue 2.0cm
{\Large Tianjun Li~\footnote{E-mail: tli@bokchoy.hep.upenn.edu,
phone: (215) 573-5820, fax: (215) 898-2010.}}
\vglue 1cm
{ Department of Physics and Astronomy \\
University of Pennsylvania, Philadelphia, PA 19104-6396 \\  
U.  S.  A.}
\end{center}

\vglue 1.5cm
\begin{abstract}
With the ansatz that there exist local or global discrete symmetries in
the 
special branes' neighborhoods, we can construct the extra dimension models
 with only zero modes, or the models which have
 large extra dimensions and arbitrarily heavy KK modes because there is
no simple relation between the
mass scales of extra dimensions and the masses of KK states.
In addition, the bulk gauge symmetry and supersymmetry can be broken on
the
special branes for all the modes, and in the bulk for the zero modes by
local and global discrete symmetries. To be explicit, we discuss the 
supersymmetric $SU(5)$ model on $M^4\times S^1/Z_2$ in which
there is a local $Z_2'$ symmetry in the special 3-brane 
neighborhood along the fifth dimension.
\\[1ex]
\\[1ex]

\end{abstract}

\vspace{0.5cm}
\begin{flushleft}
\baselineskip=12pt
August 2001\\
\end{flushleft}
\newpage
\setcounter{page}{1}
\pagestyle{plain}
\baselineskip=14pt

\section{Introduction}
Large extra dimension scenarios with branes have been an very interesting
subject
for the past few years, where the gauge hierarchy problem can be solved
because the 
physical volume of extra dimensions may be very large and 
the higher dimensional Planck scale might be low~\cite{AADD}, or the
metric for
the extra dimensions has warp factor~\cite{LRRS}. Naively, one might think
the 
masses of KK states are $\sqrt {\sum_i n_i^2/R_i^2}$ where $R_i$ is the
radius of the
$i-th$ extra dimension. However, it is shown that this is not true if one
considered
the shape moduli, and it may be possible to maintain the ratio (hierarchy)
between the higher
dimensional Planck scale and 4-dimensional Planck scale while
simultaneously 
making the KK states arbitrarily heavy~\cite{KRD}. So, a lot of
experimental bounds on the
theories with large extra dimensions are relaxed.

In this letter, we present another scenario where the higher dimensional
Planck
scale can be low while all the KK modes are projected out or
the masses of KK states can be set arbitrarily heavy.
Our ansatz is that there exist local or global discrete symmetries in the 
special branes' neighborhoods,
which become additional constraints on the KK states. 
The KK states, which satisfy
the local and global discrete
symmetries, remain in the theory, while the KK states, which do
not satisfy the local or global discrete symmetries, are projected out.
Therefore, we can construct the theories with only zero modes for all the
KK modes are projected out, or the theories which have large extra
dimensions
and arbitrarily heavy KK modes because there is no simple relation between
the
mass scales of extra dimensions and the masses of KK states.
In addition, the bulk gauge symmetry and supersymmetry can be broken on
the
special branes for the zero and KK modes, and in the bulk for the zero
modes by
local and global discrete symmetries. As an example,
we discuss the supersymmetric 
$SU(5)$ model on the space-time $M^4\times S^1/Z_2$ in which
there is a special 3-brane along the fifth dimension. In the neighborhood
of 
 the special 3-brane, there exists a local $Z_2'$ symmetry, 
which is broken globally due to
the presences of two boundary 3-branes. The bulk 4-dimensional $N=2$
supersymmetry and 
$SU(5)$ gauge symmetry are broken down to the 4-dimensional $N=1$
supersymmetry and
$SU(3)\times SU(2)\times U(1)$ gauge symmetry on the special brane for all
the states, and
in the bulk for the zero modes.
 Moreover,  all the KK states
can be projected out or can be set arbitrarily heavy although the physical
size
of the fifth dimension can be large, even at millimeter range.

\section{Local Discrete Symmetry in the Brane Neighborhood}
We assume that in a (4+n)-dimensional space-time manifold $M^4\times M^n$
where 
$M^4$ is the 4-dimensional Minkowski space-time and $M^n$ is the manifold
for extra space dimensions, there exist some topological defects, or
we call them branes for simplicity.
The special branes, which we are interested in, have co-dimension one or
more than
one. Assuming we have $K$ special branes and using $I-th$ special brane as
a representative,
 our ansatz is that in the open neighborhood
 $M^4\times U_I$ ($U_I$ $\subset $ $M^n$)
of the $I-th$ special brane, there is a local discete
symmetry\footnote{Global discrete symmetry is a
``special'' case of local discrete symmetry. The key difference is that,
the 
space-time manifold can modulo the global discrete symmetry and become a
quotient space-time manifold or orbifold.}, which forms a discrete
group $\Gamma_I$ where $I=1, 2, ...K$. And the Lagrangian is invariant
under the
local discrete symmetries. Assume the local coordinates for extra
dimensions in
the $I-th$ special brane neighborhood are $y^1$, $y^2$,
..., $y^n$, the action of any element 
$\gamma_i^I$ $\subset$ $\Gamma_I$ on $U_I$ can be expressed as
\begin{eqnarray}
\gamma_i^I: ~~~( y^1, y^2, ..., y^n) \subset U_I \longrightarrow
(\gamma_i^I y^1, 
\gamma_i^I y^2, ..., \gamma_i^I y^n)
\subset U_I ~,~\,
\end{eqnarray}
where the $I-th$ special brane position is the only fixed point, line, or
hypersurface for the
whole group $\Gamma_I$ as long as the neighborhood is small enough. 

The Lagrangian is invariant under the discrete symmetry in
the neighborhood $M^4\times U_I$ of the $I-th$ special brane, i. e., for
any element
$\gamma_i^I$ $\subset$ $\Gamma_I$
\begin{eqnarray}
{\cal L} (x^{\mu}, \gamma_i^I y^1, \gamma_i^I y^2, ..., \gamma_i^I y^n)
={\cal L} (x^{\mu},  y^1, y^2, ..., y^n) ~,~\,
\end{eqnarray}
where $ (y^1, y^2, ..., y^n) \subset U_I$. So, for a generic bulk
multiplet $\Phi$
which fills a representation of the bulk gauge group $G$, we have
\begin{eqnarray}
\Phi (x^{\mu}, \gamma_i^I y^1, \gamma_i^I y^2, ..., \gamma_i^I y^n) =
\eta_{\Phi}^I
(R_{\gamma_i^I})^{l_\Phi} \Phi (x^{\mu},  y^1, y^2, ..., y^n) 
(R_{\gamma_i^I}^{-1})^{m_\Phi}
~,~\,
\end{eqnarray} 
where $\eta^I_{\Phi}$ can be determined from the Lagrangian 
(up to $\pm1$ for the matter fields), 
$\l_{\Phi}$ and $m_{\Phi}$ are the non-negative integers
determined by the representation of $\Phi$
under the gauge group $G$.
In general, $\eta_{\Phi}^I$ is an element in $\Gamma_I$,
for example $\Gamma_I=Z_2$, $\eta_{\Phi}^I =\pm 1$. Moreover,
$R_{\gamma_i^I}$ is
an element in $G$, and $R_{\Gamma_I}$ is a discrete subgroup of $G$. We
will 
choose $R_{\gamma_i^I}$ as 
the matrix representation for $\gamma_i^I$ in the adjoint representation
of the
gauge group $G$. The consistent condition for $R_{\gamma_i^I}$ is
\begin{eqnarray}
R_{\gamma_i^I} R_{\gamma_j^I} = R_{\gamma_i^I \gamma_j^I} ~,~ 
\forall \gamma_i^I,~\gamma_j^I \subset \Gamma_I ~.~\,
\end{eqnarray} 
Mathematical speaking, the map $R:~ \Gamma_I \longrightarrow R_{\Gamma_I}
\subset G $ is
a homomorphism. 
Because the special branes are fixed under the local discrete symmetry
transformations,
the gauge group on the $I-th$ special brane is the subgroup of $G$ which
commutes with
$R_{\Gamma_I}$, {\it i. e.}, $[R_{\Gamma_I}, G]$. 
And for the zero modes, the bulk gauge group is broken down to
 the subgroup of $G$ which commutes with all $R_{\Gamma_I}$, i. e.,
$R_{\Gamma_1}$,
$R_{\Gamma_2}$, ..., $R_{\Gamma_K}$, which is important if we
wanted to discuss high rank GUT symmetry breaking. In addition, if the
theory is supersymmetric,
the special branes will preserve part of the bulk supersymmetry, and the
zero modes in the bulk also preserve part of the supersymmetry, in other
words, the
supersymmetry can be broken on the special branes for all the modes,
 and in the bulk for the zero modes.

In addition, we only have the KK states which satisfy the local
 and global discrete  symmetries in the theories because the KK modes, 
which do not satisfy the local
and global discrete symmetries, are projected out under our ansatz.
Therefore, we can construct the theories with only zero modes because all
the
KK modes are projected out, or the theories which have large extra
dimensions
and arbitrarily heavy KK states for there is no simple relation between
the
mass scales of extra dimensions and the masses of KK states. To be
explicit,
we would like to discuss the KK mode expansions in a simple scenario.

Let us consider 
the 5-dimensional space-time which can be factorized into a product of the 
ordinary 4-dimensional Minkowski space-time $M^4$ and the orbifold 
$S^1/Z_2$. The corresponding coordinates are $x^{\mu}$, ($\mu = 0, 1, 2,
3$),
$y\equiv x^5$, and the radius for the fifth dimension is $R$.
The orbifold $S^1/Z_2$ is obtained by $S^1$ moduloing the equivalent class
$y \sim -y$, and then, there are two
fixed points: $y=0$ and $y=\pi R$. Moreover, we assume that there is a 
special 3-brane along the fifth dimension, which is located at 
$ 0< y=s <\pi R$. Our ansatz is that, there is a local $Z_2'$
symmetry in the neighborhood of the special 3-brane, which becomes a
global
$Z_2'$ symmetry if $s=\pi R/2$. Mathematical speaking, we define
$y'\equiv y-s$, and in the open neighborhood of the sepcial 3-brane, 
we have the equivalent class $y' \sim -y'$, which can not be moduloed
because
it is not a global symmetry. Essentially speaking, we consider the
5-dimensional
space-time $M^4\times S^1$ with four special
3-branes which are located at $y=0, \pi R, -s, s$,
and the system has one non-equivalent global $Z_2$ symmetry and one 
non-equivalent local $Z_2'$ symmetry.

For a generic bulk multiplet $\Phi(x^{\mu}, y)$ which fills a
representation of the gauge
group $G$,
we can define two parity operators $P$ and $P'$ for the $Z_2$ and
$Z_2'$ symmetries, respectively
\begin{eqnarray}
\Phi(x^{\mu},y)&\to \Phi(x^{\mu},-y )=\eta_{\Phi}
P^{\l_{\Phi}}\Phi(x^{\mu},y)
(P^{-1})^{m_{\Phi}}~,~\,
\end{eqnarray}
\begin{eqnarray}
\Phi(x^{\mu},y')&\to \Phi(x^{\mu},-y' )= \eta'_{\Phi} (P')^{\l_{\Phi}}
\Phi(x^{\mu},y')
(P^{'-1})^{m_{\Phi}}
 ~,~\,
\end{eqnarray}
where $\eta_{\Phi} = \pm 1$, $\eta'_{\Phi} = \pm 1$. And in the
discussions of next section,
for simplicity, we assume that $\eta_{\Phi} = \eta'_{\Phi}$.

Denoting the field $\phi$ with ($P$, $P'$)=($\pm, \pm$) by $\phi_{\pm
\pm}$, we obtain that
if $s/(\pi R-s)$ is not a rational number, then, we only have the zero
modes because
all the KK modes are projected out.
And if $s/(\pi R-s)$ is a rational number, we can define the relatively
prime integers 
$p$ and $q$ by 
\begin{eqnarray}
{p\over q}={s\over {\pi R-s}} ~.~\,
\end{eqnarray}
If $p$ or $q$ is even, the fields $ \phi_{+-} (x^\mu, y)$
and $\phi_{-+} (x^\mu, y)$ are projected out. Only when $p$ and $q$ are
odd, we
have the full KK mode expansions 
\begin{eqnarray}
  \phi_{++} (x^\mu, y) &=& 
      \sum_{n=0}^{\infty} \frac{1}{\sqrt{2^{\delta_{n,0}} \pi R}} 
      \phi^{(2n)}_{++}(x^\mu) \cos{2ny \over r}~,~\,
\end{eqnarray}
\begin{eqnarray}
  \phi_{+-} (x^\mu, y) &=& 
      \sum_{n=0}^{\infty} \frac{1}{\sqrt{\pi R}} 
      \phi^{(2n+1)}_{+-}(x^\mu) \cos{(2n+1)y \over r}~,~\,
\end{eqnarray}
\begin{eqnarray}
  \phi_{-+} (x^\mu, y) &=& 
      \sum_{n=0}^{\infty} \frac{1}{\sqrt{\pi R}} \,
      \phi^{(2n+1)}_{-+}(x^\mu) \sin{(2n+1)y \over r}~,~\,
\end{eqnarray}
\begin{eqnarray}
  \phi_{--} (x^\mu, y) &=& 
      \sum_{n=0}^{\infty} \frac{1}{\sqrt{\pi R}} 
      \phi^{(2n+2)}_{--}(x^\mu) \sin{(2n+2)y \over r}~,~\,
\end{eqnarray}
where 
\begin{eqnarray}
r={{2R}\over {p+q}}~,~\,
\end{eqnarray}
and $n$ is a non-negative integer.
The 4-dimensional fields $\phi^{(2n)}_{++}$, $\phi^{(2n+1)}_{+-}$, 
$\phi^{(2n+1)}_{-+}$ and $\phi^{(2n+2)}_{--}$ acquire masses 
$2n/r$, $(2n+1)/r$, $(2n+1)/r$ and $(2n+2)/r$ upon the compactification.
Zero modes are contained only in $\phi_{++}$ fields,
thus, the matter content of massless sector is smaller
than that of the full 5-dimensional multiplet.
Moreover, only $\phi_{++}$ and $\phi_{+-}$ fields have non-zero
values at $y=0$ and $y=\pi R$, and only $\phi_{++}$ and $\phi_{-+}$
 fields have non-zero values at $y=s$.
By the way, when $p=q=1$, i. e., $s=\pi R/2$, we obtain the previous 
KK mode expansions~\cite{Nbhn1}.

In short, from our simple scenario, we obtain that: (I) 
if $s/(\pi R-s)$ is not a rational number, we only have zero modes, i. e.,
$\phi_{++}^{(0)}(x^{\mu})$;
(II) if $p$ or $q$ is even, we only have the fields
$\phi_{++}^{(2n)}(x^{\mu})$
and $\phi_{--}^{(2n+2)}(x^{\mu})$;
(III) if $p$ and $q$ are odd, we will have the KK mode expansions
for all the fields $\phi^{(2n)}_{++}(x^\mu)$,
$\phi^{(2n+1)}_{+-}(x^\mu)$, $\phi^{(2n+1)}_{-+}(x^\mu)$ and 
$\phi^{(2n+2)}_{--}(x^\mu)$. In addition, because $0 < r \le R$,
 the masses of KK states ($n/r$)
can be set arbitrarily heavy if we choose suitable $p$ and $q$, for
instance, if $1/R$ is about TeV, $p=10^{13}-1$ and $q=10^{13}+1$,
or $p=10^{13}+1$ and $q=1$,
we obtain that $1/r$ is about $10^{16}$ GeV, which is the usual GUT scale.
Therefore, there is no simple relation between the physical size of 
the fifth dimension and the mass scales of KK modes. Furthermore,
the gauge symmetry and supersymmetry can be broken on the special 
3-brane for all the modes,
and in the bulk for the zero modes by choosing suitable $P'$.

\section{Application to the Supersymmetric $SU(5)$ Model on $M^4\times
S^1/Z_2$}
We would like to discuss the supersymmetric $SU(5)$ model on the
space-time 
$M^4\times S^1/Z_2$ with a special 3-brane which
has local $Z_2'$ symmetry along the fifth dimension.
We assume that the $SU(5)$ gauge fields and two 5-plet Higgs
hypermultiplets
in the bulk, and the Standard Model fermions can be on the 3-brane or in
the
bulk. The $SU(5)$ gauge symmetry breaking mechanism is similar to those 
discussed by a lot of papers recently~\cite{SBPL, HN}.

As we know, the $N=1$ supersymmetric theory in 5-dimension have 8 real
supercharges,
corresponding to $N=2$ supersymmetry in 4-dimension. The vector multiplet
physically contains a vector boson $A_M$ where $M=0, 1, 2, 3, 5$, 
two Weyl gauginos $\lambda_{1,2}$, and a real scalar $\sigma$. 
In terms of 4-dimensional $N=1$ supersymmetry
language, it contains a vector multiplet $V(A_{\mu}, \lambda_1)$ and
a chiral multiplet $\Sigma((\sigma+iA_5)/\sqrt 2, \lambda_2)$ which
transform
in the adjoint representation of $SU(5)$.
And the 5-dimensional hypermultiplet physically has two complex scalars
$\phi$ and $\phi^c$, a Dirac fermion $\Psi$, and can be decomposed into 
two chiral mupltiplets $\Phi(\phi, \psi \equiv \Psi_R)$
and $\Phi^c(\phi^c, \psi^c \equiv \Psi_L)$, which transform as conjugate
representations
of each other under the gauge group. For instance,
 we have two Higgs chiral multiplets $H_u$ and $H_d$, which transform 
as $ 5$ and $\bar 5$ under SU(5) gauge symmetry, and their
mirror $H_u^c$ and $H_d^c$, which transform as $ \bar 5$ and $ 5$ under
SU(5) gauge symmetry.

The general action for the $SU(5)$ gauge fields and their couplings to the
bulk hypermultiplet $\Phi$ is~\cite{NAHGW} 
\begin{eqnarray}
S&=&\int{d^5x}\frac{1}{k g^2}
{\rm Tr}\left[\frac{1}{4}\int{d^2\theta} \left(W^\alpha W_\alpha+{\rm H.
C.}\right)
\right.\nonumber\\&&\left.
+\int{d^4\theta}\left((\sqrt{2}\partial_5+ {\bar \Sigma })
e^{-V}(-\sqrt{2}\partial_5+\Sigma )e^V+
\partial_5 e^{-V}\partial_5 e^V\right)\right]
\nonumber\\&&
+\int{d^5x} \left[ \int{d^4\theta} \left( {\Phi}^c e^V {\bar \Phi}^c +
 {\bar \Phi} e^{-V} \Phi \right)
\right.\nonumber\\&&\left.
+ \int{d^2\theta} \left( {\Phi}^c (\partial_5 -{1\over {\sqrt 2}} \Sigma)
\Phi + {\rm H. C.}
\right)\right]~.~\,
\end{eqnarray}

Because the action is invariant under the parities
 $P$ globally and $P'$ in the special 3-brane neighbohood,
 we obtain that under the parity operator $P$, the
vector multiplet transforms as
\begin{eqnarray}
V(x^{\mu},y)&\to  V(x^{\mu},-y) = P V(x^{\mu}, y) P^{-1}
~,~\,
\end{eqnarray}
\begin{eqnarray}
 \Sigma(x^{\mu},y) &\to\Sigma(x^{\mu},-y) = - P \Sigma(x^{\mu}, y) P^{-1}
~,~\,
\end{eqnarray}
if the hypermultiplet $\Phi$ is a $5$ or $\bar 5$ $SU(5)$ multiplet, we
have 
\begin{eqnarray}
\Phi(x^{\mu},y)&\to \Phi(x^{\mu}, -y)  = \eta_{\Phi} P \Phi(x^{\mu},y)
~,~\,
\end{eqnarray}
\begin{eqnarray}
\Phi^c(x^{\mu},y) &\to \Phi^c(x^{\mu}, -y)  = -\eta_{\Phi} P
\Phi^c(x^{\mu},y)
~,~\,
\end{eqnarray}
and if the hypermultiplet $\Phi$ is a $10$ or $\bar {10}$ $SU(5)$
multiplet, we have
\begin{eqnarray}
\Phi(x^{\mu},y)&\to \Phi(x^{\mu}, -y)  = \eta_{\Phi} P \Phi(x^{\mu},y)
P^{-1}
~,~\,
\end{eqnarray}
\begin{eqnarray}
\Phi^c(x^{\mu},y) &\to \Phi^c(x^{\mu}, -y)  = -\eta_{\Phi} P
\Phi^c(x^{\mu},y) P^{-1}
~,~\,
\end{eqnarray}
where $\eta_{\Phi} = \pm 1$. And the transfomations
of vector multiplet and hypermultiplet under the local $Z_2'$ symmetry
$P'$
are similar to those under the global $Z_2$ symmetry $P$.

We choose the following matrix representations for the global parity $P$
and 
local parity $P'$ which are expressed in the adjoint representaion of 
$SU(5)$
\begin{equation}
P={\rm diag}(+1, +1, +1, +1, +1)~,~P'={\rm diag}(+1, +1, +1, -1, -1)
 ~.~\,
\end{equation}
So, upon the local $P'$ parity,
 the gauge generators $T^A$ where A=1, 2, ..., 24 for $SU(5)$
are separated into two sets: $T^a$ are the gauge generators for
the Standard Model gauge group, and $T^{\hat a}$ are the other broken
gauge generators 
\begin{equation}
P~T^a~P^{-1}= T^a ~,~ P~T^{\hat a}~P^{-1}= T^{\hat a}
~,~\,
\end{equation}
\begin{equation}
P'~T^a~P^{'-1}= T^a ~,~ P'~T^{\hat a}~P^{'-1}= - T^{\hat a}
~.~\,
\end{equation}

Choosing $\eta_{H_u}=+1$ and $\eta_{H_d}=+1$, we obtain the 
 particle spectra, which are given in 
Table 1. And the gauge fields, Higgs fields, and gauge groups on the
3-branes
are given in Table 2.  The bulk 4-dimensional $N=2$ supersymmetry and 
$SU(5)$ gauge symmetry are broken down to the 4-dimensional $N=1$
supersymmetry and
$SU(3)\times SU(2)\times U(1)$ gauge symmetry on the special 3-brane for
all the states, and
in the bulk for the zero modes.
Including the KK states, the gauge symmetry in the bulk and
 on the boundary 3-brane which is located at $y=0$ or $y=\pi R$ is
$SU(5)$. By the way,
the 4-dimensional supersymmetry on the 3-branes at 
$y=0, s, \pi R$ is ${1/2}$ of the bulk 4-dimensional supersymmetry or
$N=1$ due
to the $Z_2$ or $Z_2'$ symmetry.

\renewcommand{\arraystretch}{1.4}
\begin{table}[t]
\caption{Parity assignment and masses ($n\ge 0$) of the fields in the
SU(5) 
 gauge and Higgs multiplets.
The indices $F$, $T$ are for doublet and triplet, respectively. 
\label{tab:chiral}}
\vspace{0.4cm}
\begin{center}
\begin{tabular}{|c|c|c|}
\hline        
$(P,P')$ & field & mass\\ 
\hline
$(+,+)$ &  $V^a_{\mu}$, $H^F_u$, $H^F_d$ & ${{2n}\over r}$ \\
\hline
$(+,-)$ &  $V^{\hat{a}}_{\mu}$,  $H^T_u$, $H^T_d$ & ${{2n+1}\over r}$  \\
\hline
$(-,+)$ &  $\Sigma^{\hat{a}}$, ${H}^{cT}_u$, ${H}^{cT}_d$  & ${{2n+1}\over
r}$ \\
\hline
$(-,-)$ &  $\Sigma^a$, ${ H}^{cF}_u$, ${H}^{cF}_d$ &  ${{2n+2}\over r}$\\
\hline
\end{tabular}
\end{center}
\end{table}

\renewcommand{\arraystretch}{1.4}
\begin{table}[t]
\caption{The gauge fields, Higgs fields and gauge group on
 the 3-branes which are located
at  $y=0$, $y=s$, and  $y=\pi R$.
\label{tab:chiralII}}
\vspace{0.4cm}
\begin{center}
\begin{tabular}{|c|c|c|}
\hline        
Brane position & field & gauge group\\ 
\hline
$y=0$ & $V_{\mu}^A$, $H_u$, $H_d$ & $SU(5)$  \\
\hline
$y=s$ &  $V_{\mu}^a$, $\Sigma^{\hat a}$, $H_u^F$, $H_d^F$, 
${ H}_u^{cT}$, ${ H}_d^{cT}$ & $SU(3)\times SU(2) \times U(1)$\\
\hline
$y=\pi R$ &   $V_{\mu}^A$, $H_u$, $H_d$ & $SU(5)$ \\
\hline
\end{tabular}
\end{center}
\end{table}

\subsection{The Standard Model Fermions on the 3-Brane} 
If the Standard Model fermions were on the special 3-brane at $y=s$,
the gauge symmetry is $SU(3)\times SU(2)\times U(1)$. Because the Higgs
triplets are projected out on the special 3-brane, the Yukawa couplings 
are no more restricted than in the usual
4-dimensional Minimal Supersymmetric Standard Model, 
and there is enough flexibility to accommodate fermion
masses and mixings. In short, the Yukawa terms in the superpotential
of the 5-dimensional effective Lagrangian are
\begin{eqnarray}
W_{\rm Yukawa}= \int d^2\theta \delta(y-s)\sum_{I=1}^3 \sum_{J=1}^3 \left(
h^u_{IJ} Q^I H_u^F {\bar U}^J +h^d_{IJ} Q^I H_d^F {\bar D}^J
+ h^l_{IJ} L^I H_d^F {\bar E}^J \right)~,~
\end{eqnarray}
where the $Q$, $U$, $D$, $L$, $E$ denote
the quark $SU(2)_L$ doublet, right-handed up-type quark, 
right-handed down-type quark, lepton/neutrino
$SU(2)_L$ doublet, and right-handed lepton, respectively.
So, we avoid the wrong 
 prediction of the first and second generation
fermion mass ratios, $m_d/m_e$ and $m_s/m_{\mu}$ in the usual 
4-dimensional $SU(5)$ model. And if 
the Standard Model fermions only preserve the
$SU(3)\times SU(2)\times U(1)$ gauge symmetry, there are no proton decay
problem at all.
However, we can not explain the charge quantization. 
By the way, we can put two Higgs doublets on the special 3-brane instead
of putting
two Higgs 5-plets in the bulk. 

If the Standard Model fermions were on the boundary 3-brane at $y=0$ or
$y=\pi R$, 
the gauge symmetry
is $SU(5)$.
However, we can not define the proper Yukawa terms in the superpotential
unless
$s=\pi R/2$. So, we will not discuss it here. 

\subsection{The Standard Model Fermions in the Bulk} 
Now, we consider the scenario where the Standard Model fermions 
are in the Bulk. We will double the generations in the bulk due to
the $P'$ projections, i. e., we will
have $T^I + {\bar F}^I + {T}^{cI} +  {\bar F}^{cI}$ and
$T^{'I} + {\bar F}^{'I} + {T}^{'cI} + {\bar F}^{'cI}$ where $I=1, 2, 3$,
 $c$ denotes the charge conjugation, and $\eta_{T^I}=\eta_{{\bar
F}^{I}}=1$,
$\eta_{T^{'I}}=\eta_{{\bar F}^{'I}}=-1$.
And each generation in the Standard Model comes from the zero modes of
two generations. Moreover,
the particle spectra for the bulk fermions are given in Table 3.

\renewcommand{\arraystretch}{1.4}
\begin{table}[t]
\caption{Parity assignment and masses ($n\ge 0$) of the bulk fermions in
the 
 $10 + \bar 5$ $SU(5)$  multiplets. 
\label{tab:chiralA}}
\vspace{0.4cm}
\begin{center}
\begin{tabular}{|c|c|c|}
\hline        
$(P,P')$ & field & mass\\ 
\hline
$(+,+)$ & $T^I_{\bar U}$, $T^I_{\bar E}$, ${\bar F}^I_{L}$, $T^{'I}_{Q}$, 
${\bar F}^{'I}_{\bar D} $ & ${{2n}\over r}$ \\
\hline
$(+,-)$ & $T^I_{Q}$, ${\bar F}^{I}_{\bar D} $,
$T^{'I}_{\bar U}$, $T^{'I}_{\bar E}$, ${\bar F}^{'I}_{L}$ & ${{2n+1}\over
r}$ \\
\hline
$(-,+)$ &  $T^{cI}_{\bar Q}$, ${\bar F}^{cI}_{ D} $,
$T^{'cI}_{ U}$, $T^{'cI}_{ E}$, ${\bar F}^{'cI}_{\bar L}$ & ${{2n+1}\over
r}$ \\
\hline
$(-,-)$ &   $T^{cI}_{ U}$, $T^{cI}_{ E}$, ${\bar F}^{cI}_{\bar L}$,
$T^{'cI}_{\bar Q}$, 
${\bar F}^{'cI}_{ D} $ &  ${{2n+2}\over r}$\\
\hline
\end{tabular}
\end{center}
\end{table}

Because the fermions and  Higgs fields are in the bulk,
we need to define the bulk Yukawa terms. The relevant bulk Yukawa terms in
the
superpotential are
\begin{eqnarray}
W_{\rm Yukawa} &=&
\int d^2\theta \sum_{I=1}^3 \sum_{J=1}^3 \left( (\theta(y)-\theta(-y))
h^u_{IJ} T^{'I} T^J H_u
\right.\nonumber\\&&\left.
+ h^d_{IJ} T^{'I} {\bar F}^{'J} H_d+
h^l_{IJ} T^{I} {\bar F}^{J} H_d \right)~,~\,
\end{eqnarray}
where $\theta(x)$ is the step function defined as $\theta(y) =1$ for $y\ge
0$,
and $\theta(y)=0$ for $y<0$.
And for the zero modes, the Yukawa terms in the superpotential become
\begin{eqnarray}
W_{\rm Yukawa} = 
\int d^2\theta \sum_{I=1}^3 \sum_{J=1}^3 \left(h^u_{IJ} T^{'I}_Q T^J_{\bar
U} H_u^F+
h^d_{IJ} T^{'I}_Q {\bar F}^{'J}_{\bar D} H_d^F+
h^l_{IJ} T^{I}_{\bar E} {\bar F}^{J}_{L} H_d^F \right)~.~\,
\end{eqnarray}
In this scenario, we avoid the wrong $SU(5)$ prediction of the first and
second generation
fermion mass ratios, and have charge quantization. In addition, the
tree-level
proton decays by exhange $X$, $Y$ and Higgs triplets are absent because
$D$ and $L$,
$Q$ and $U/E$ come from different hypermultiplets.

\subsection{Phenomenology Comments}
In our model, the physical radius of the fifth dimension is $R$, while the
masses of KK
states is $n/r$. Because $0 < r \le R$,  we can have large extra dimension
and arbitrarily
heavy KK states. For example, $1/R$ might be at TeV scale and the masses
of KK states
can be at $10^{16}$ GeV. Especially, when we consider the
non-supersymmetric $SU(5)$ model on the space-time $M^4\times S^1/Z_2$, 
we might solve the gauge
hierarchy problem by having large extra dimension, and avoid the proton
decay by
 giving $X$, $Y$, and Higgs triplet masses about $10^{16}$ GeV because
they do not
have zero modes. Furthermore, we can push the physical size of the fifth
dimension to
the millimeter range, while the masses of KK states are still at
 TeV scale. 

In addition, if one considered
the Scherk-Schwartz mechanism of
 supersymmetry breaking, the soft mass scale and the possible $\mu$
term might be at order of $1/R$, which can be at TeV scale naturally,
unlike in 
previous models, where $1/R$ is the GUT scale~\cite{SBPL, HN}.

\section{Discussion and Conclusion}
In section 3, one can also choose the matrix representations for the
global parity $P$ and 
local parity $P'$ as
\begin{equation}
P={\rm diag}(+1, +1, +1, -1, -1)~,~P'={\rm diag}(+1, +1, +1, +1, +1)
 ~.~\,
\end{equation}
So, the gauge group on the boundary 3-brane at $y=0$ or $y=\pi R$ is
$SU(3)\times SU(2) \times U(1)$, and the gauge group on the special
3-brane at $y=s$ is $SU(5)$. The discussions for the 
Standard Model fermions on the boundary 3-brane or in the bulk are similar
to those in above section. So, let us discuss the case where 
 the Standard Model fermions are on
the special 3-brane with $SU(5)$ gauge symmetry. As we know, 
 the transformation properties of the quark and lepton superfields under
the
$Z_2$ symmetry or $P$ parity are determined by the requirement that 
any operators on the special 3-brane must transform $SU(5)$ covariantly
under $Z_2$ symmetry.
Because the kinetic terms for the
$10$  $T^I$ and $\bar 5$ ${\bar F}^J$  $SU(5)$ multiplets must
transform covariantly under $Z_2$ 
symmetry  where $I, J=1, 2, 3$, there are only four possibilities for the 
assignment of the 
$P$ quantum numbers: (I) $P(Q,U,D,L,E) = \pm(+,-,-,+,-)$ or (II) 
$P(Q,U,D,L,E) = \pm(-,+,-,+,+)$. And then, 
we can obtain the transformation properties of the Yukawa 
couplings $[T^I T^J H_{u}]_{\theta^2}$ and 
$[T^I {\bar F}^J H_{d}]_{\theta^2}$ under the $P$ parity: 
for the case (I)
 $P(T^I T^J H_{u}) = 
P(T^I {\bar F}^J H_{d}) = -$;
and for the case (II) $P(T^I T^J H_{u}) = 
-P(T^I {\bar F}^J H_{d}) = -$. Therefore,
the $Z_2$ invariant Yukawa interactions in the superpotential are
\begin{eqnarray}
 W_{\rm Yukawa}  &=& \int d^2 \theta\sum_{I=1}^3 \sum_{J=1}^3 \Biggl[
    \frac{1}{2} \{ \delta(y-s) - \delta(y+s) \}
    \sqrt{2 \pi R}\, h^u_{IJ} T^{I} T^{J} H_{u} 
\nonumber\\
&& + \frac{1}{2} \{ \delta(y-s) \mp \delta(y+s) \}
    \sqrt{2 \pi R}\, h_d^{I J} T^{I} {\bar F}^{J} H_{d}
  \Biggr]~,~\,
\label{eq:5d-Yukawa}
\end{eqnarray}
where $\mp$ takes $-$ and $+$ in the case of (I) and (II), 
respectively. However, we have
 the wrong $SU(5)$ prediction of the first and second generation
fermion mass ratios, which
may be solved by adding bulk hypermultiplets which transform as 
$5 +\bar 5$~\cite{HN}.

Furthermore, one can consider the models on the space-time $M^4\times M^1$
with
two local $Z_2$ symmetries, and the models on the space-time
$M^4\times M^1\times M^1$, $M^4\times A^2$ and $M^4\times T^2$
 with various local discrete symmetries, like
$(Z_2)^3$, $(Z_2)^4$, $Z_n$, etc. Those generalizations will
be discussed elsewhere~\cite{LTJ}.

In short, with the ansatz that there exist local or global discrete
symmetries in the 
special branes' neighborhoods, we can construct the extra dimension models
 with only zero modes, or the models which have
 large extra dimensions and arbitrarily heavy KK modes because there is
no simple relation between the
mass scales of extra dimensions and the masses of KK states.
In addition, the bulk gauge symmetry and supersymmetry can be broken on
the
special branes for all the modes, and in the bulk for the zero modes by
local and global discrete symmetries. To be explicit, we discuss the 
supersymmetric $SU(5)$ model on the space-time $M^4\times S^1/Z_2$ in
which
there is a local $Z_2'$ symmetry in the special 3-brane 
neighborhood along the fifth dimension.

\section*{Acknowledgments}
This research was supported in part by the U.S.~Department of Energy under
 Grant No.~DOE-EY-76-02-3071.

\end{document}